\documentclass[prl,amsmath,twocolumn,floatfix,showpacs]{revtex4}
\usepackage{graphicx} 
\usepackage{latexsym}
\usepackage{amssymb}
\newcommand{\lo}{\left \langle}
\newcommand{\rc}{\right \rangle}

\usepackage{amssymb}
\begin{document}
\title{Can one predict DNA Transcription Start Sites by studying bubbles?}

\author{Titus S. van Erp$^{1,2}$, Santiago Cuesta-Lopez$^{2,3}$, 
Johannes-Geert 
Hagmann$^{1,2}$, and Michel Peyrard$^2$} 
\affiliation{$1$  Centre Europ\'een de Calcul Atomique et 
Mol\'eculaire (CECAM)\\
$2$ Laboratoire de Physique, Ecole Normale Sup\'erieure de Lyon,
46 all\'ee d'Italie, 69364 Lyon Cedex 07, France\\
$3$ Dept. Condensed Matter Physics and Institut of Biocomputation and 
Complex Systems. University of Zaragoza, c/ Pedro Cerbuna s/n 50009 Spain }

\begin{abstract}
It has been speculated that bubble formation of several base-pairs due to thermal fluctuations is indicatory for biological active sites. Recent evidence, based on experiments and molecular dynamics (MD) simulations using the Peyrard-Bishop-Dauxois model, seems to point in this direction. However, sufficiently large bubbles appear only seldom which makes an accurate calculation difficult even for minimal models. In this letter, we introduce a new method that is orders of magnitude faster than MD.  Using this method we show that the present evidence is unsubstantiated.
\end{abstract}


\maketitle
Double stranded DNA (dsDNA) is not a static entity.
In solution, the bonds between bases on opposite strands
can break even at room temperature. This can happen for
entire regions of the dsDNA chain, which then form
bubbles of several base-pairs (bp). These phenomena
are important
for  biological processes such as replication and transcription.
The local opening of the DNA double helix at the 
transcription start site (TSS) is 
a crucial step
for the transcription of the
genetic code.
This opening is driven by proteins
but the 
intrinsic fluctuations of DNA itself probably play an important role.
The statistical and dynamical
properties of these denaturation bubbles and their relation
to biological functions
have therefore been subject of many
experimental and theoretical studies.
It is known that the denaturation process of finite
DNA chains is not simply determined 
by the fraction of strong (GC) or weak (AT) base-pairs.  
The sequence specific order is important. Special sequences can have a high 
opening rate despite a high fraction of GC base pairs \cite{Dornberger}. 
For supercoiled DNA, it has been suggested that these sequences are related to 
places known to be important for initiating and regulating 
transcription~\cite{PNAS1}.
For dsDNA, Choi et al found 
evidence that the formation of bubbles  
is directly related the transcription sites~\cite{ChoiNuc2004}. 
In particular, their results indicated that 
the TSS could be predicted on basis of the formation probabilities 
for bubbles of ten or more base-pairs in absence of proteins. 
Hence, the secret of the TSS is not in the protein that reads the code, 
but really a characteristics of DNA as expressed by the statement:
\emph{DNA directs its own transcription}~\cite{ChoiNuc2004}. 
In that work,
S1 nuclease cleavage
experiments were compared with molecular dynamics (MD) simulations
on the Peyrard-Bishop-Dauxois (PBD) model~\cite{PB,PBD} of DNA.
The method used is not without limitations.
The
S1 nuclease cleavage is related to opening, 
but many other complicated factors are involved. 
Moreover, theoretical and computational studies have to rely on 
simplified models and considerable computational power.
As the formation of large bubbles
occurs only seldom in a microscopic system, 
MD or Monte Carlo (MC) methods
suffer from demanding computational efforts
to obtain sufficient accuracy. 
Nevertheless, the probability profile found for bubbles of ten and higher 
showed a striking correlation with the experimental results yielding 
pronounced peaks at the TSS~\cite{ChoiNuc2004}.
Still, the large statistical uncertainties 
make this correlation questionable.  
To make the assessment absolute, we would either need extensively 
long or exceedingly many 
simulation runs or a different method that is significantly 
faster than MD. 

In this letter, 
we introduce 
such a
method for the calculation of bubble statistics 
for first neighbor interaction models like the PBD. 
We applied it  to the sequences studied in Refs.~\cite{ChoiNuc2004} and,
to validate the method and to compare its efficiency, we repeated
the MD simulations with 100 times longer runs.  
The new method shows results consistent with MD but with a lot higher 
accuracy than these considerably longer simulations.
Armed with this novel method, we make a full analysis of 
preferential opening sites
for bubbles of any length.
This analysis shows that there is no strict analogy 
between these preferential sites and the TSS using equilibrium statistics.
Hence, the previously found correlation
must have been either accidental or due to some non-equilibrium effect,
which remains speculative.
We discuss this issue and, more generally, the required
theoretical and experimental advancements that could address the
title's question definitely.

The PBD model reduces the myriad degrees of freedom of DNA
to an one-dimensional chain of effective atom compounds 
describing the relative base-pair 
separations $y_i$ from the ground state positions.
The total potential energy $U$ for an $N$ base-pair DNA chain is then given by 
$U(y^N)=V_1(y_1)+\sum_{i=2}^N V_i(y_i) +  W(y_i,y_{i-1})$
with $y^N\equiv \{y_i \}$ the set of relative base pair positions and
\begin{eqnarray}
V_i(y_i) &=& D_i \Big( e^{-a_i y_i}-1\Big)^2  \\
W(y_i,y_{i-1}) &=& \frac{1}{2} K \Big( 1+\rho e^{-\alpha(y_i+y_{i-1})}\Big)(y_i
- y_{i-1})^2 \nonumber
\label{PBpot}
\end{eqnarray}
The first term $V_i$ is the on site Morse potential describing the 
hydrogen bond interaction between bases on opposite strands. 
$D_i$ and $a_i$ determine the depth and width of the Morse potential  
and are different for the  AT and GC base-pair.
The stacking potential $W$ 
consists of a harmonic and a nonlinear term. 
The second term was later introduced~\cite{PBD} and mimics the effect
of decreasing overlap between  $\pi$
electrons when one of two neighboring base move out of stack.
As a result, the effective coupling constant of the stacking interaction 
drops from $K'=K(1+\rho)$ down to $K'=K$.
It is due to this term that the observed 
sharp phase transition in denaturation   
experiments
can be reproduced. 
All interactions with the solvent and the ions 
are effectively included in the force-field. The constants
$K, \rho,\alpha, D_{\rm AT}, D_{\rm GC}, a_{\rm AT}, a_{\rm GC}$ 
were parameterized in Ref.~\cite{CAGI} and tested on
denaturation curves of short heterogeneous  DNA segments. 
These examples show that, despite its simplified character, the model is able 
to give a quantitative description of DNA. Most importantly, it allows  
to study the statistical and dynamical behavior 
of very long heterogeneous DNA sequences, 
which is impossible for any atomistic 
model.

Despite these successes, 
it is important to realize the limitations of the model. 
The PBD model treats the A and T bases and the G and C bases as 
identical objects. The stacking interaction is also independent of the nature 
of the bases.
Moreover, there is a subtle point that needs further explanation.
As the PBD model basically represents
a single dsDNA in an infinite solution, the probability for
complete denaturation of a molecule of finite length, 
resulting in two single stranded DNAs,
tends to unity with increasing time at \emph{any temperature}.
In the experiments, where the amount of solvated DNA is not 
infinitely diluted,  this effect is counterbalanced by the recombination 
mechanism where two single stranded chains in solution come together and match 
their complementary bases. 
Hence, in our calculations we will restrict 
the configurational space to the dsDNA only, 
first of all because it is a very good 
approximation in comparison to experiments which are not performed
in the immediate vicinity of the denaturation transition and, secondly, 
because it is a necessary condition to give a relevant meaning
to the ensemble averages calculated within the PBD model.

In microscopic terms, a configuration $y^N$ is called a dsDNA molecule 
when $y_i < y_0$ for at least one $i \in [1:N]$ with $y_0$ the opening
threshold definition.
Similarly, a configuration is completely denaturated whenever 
$y_i > y_0$ for all $i$.
The statistical average $\lo A(y^N) \rc$ 
is equivalent to the ratio of two $N$-dimensional integrals
$\lo A \rc =   \int \mathrm{d}y^N A(y^N) \varrho(y^N)/ 
\int \mathrm{d}y^N  \varrho(y^N)$ with $\mathrm{d}y^N \equiv \mathrm{d}y_N
\mathrm{d}y_{N-1} \ldots  \mathrm{d}y_1$ and $\varrho$ the 
probability distribution
density. Numerical integration calculates these integrals explicitly, while
MD and MC calculates only the ratio.
Usually, the dimensionality of the system prohibits
direct numerical integration making MD and MC far favorable.
However, an increase of the computational efforts by a factor of two
reduces the error by only a factor of $\sqrt{2}$ in MD and MC, while
the reduction can be quite dramatic in low dimensional systems using 
numerical integration. 
In the following, 
we show how to exploit this
by creating 
an effective reduction of the 
dimensions
yielding an orders-of-magnitude faster algorithm for the bubble statistics 
calculation.
To explain the algorithm, we need to define a set of functions
\begin{eqnarray}
\theta_i(y_i)=\theta(y_i-y_0), \qquad  \bar{\theta}_i(y_i)=\theta(y_0-y_i)
\end{eqnarray}
 where $\theta(\cdot)$ equals the Heaviside step function. $\theta_i$ equals 1
if the base-pair is open and is zero otherwise.
$\bar{\theta}_i$ is the reverse. 
These functions
indicate whether a base-pair is open or closed. Using these, we define
\begin{eqnarray}
\theta_i^{[m]} &\equiv& \bar{\theta}_{i-\frac{m}{2}}  
\bar{\theta}_{i+\frac{m}{2
}+1} \prod_{j=i-\frac{m}{2}+1}^{i+\frac{m}{2}} \theta_j \textrm{ for $m$ even} 
\nonumber \\
&\equiv& \bar{\theta}_{i-\frac{m+1}{2}}  
\bar{\theta}_{i+\frac{m+1}{2}} \prod_{j
=i-\frac{m-1}{2}}^{i+\frac{m-1}{2}} \theta_j \textrm{ for $m$ odd}
\end{eqnarray}
which are 1 (0 otherwise) if and only if $i$ is at the center of a bubble that 
has exactly size $m$. To shorten the notation we have dropped the $y_i$ 
dependencies.
For even numbers it is a bit arbitrary where to place the center, 
but we defined it as the base
directly to the left of the midpoint of the bubble. In order to have these 
quantities defined also near
the ends of the chain, we use $\bar{\theta}_i=1$ for $i = 0$ and $i=N+1$.
The properties of interest are the probabilities
for bubbles of size $m$ centered at base-pair $i$ provided that the molecule 
is in the double stranded configuration.
\begin{eqnarray}
\lo \theta_i^{[m]}\rc_\mu &\equiv& 
\frac{\lo \theta_i^{[m]} \mu \rc }{\lo \mu \rc} \quad \textrm{ with } \quad
\mu  =1-\prod_{i=1}^N \theta_i \nonumber \\
&\equiv & \frac{Z_{\theta_i^{[m]}} }{Z-Z_\Pi}
\end{eqnarray}
Here $\mu=1$ except when all bases are open; then $\mu=0$. 
The partition function integrals are given by:
\begin{align}
&Z = \int  {\mathrm d} y^N e^{-\beta U(y^N)}, \quad 
Z_{\theta_i^{[m]}} = \int {\mathrm d} y^N e^{-\beta U(y^N)}
\theta_i^{[m]} \nonumber \\
&Z_\Pi =  \int  {\mathrm d} y^N e^{-\beta U(y^N)} \times \prod_j
\theta_j.  
\end{align}
Note that both $Z$ as $Z_\Pi$ are infinite, but their difference is  
well defined.
Now, we can make use of the fact that all integrals $Z_X$ are of the 
factorizable form
$Z_X=\int \mathrm{d} y^N a_X^{(N)}(y_N,y_{N-1})
\ldots a_X^{(3)}(y_{3},y_{2}) a_X^{(2)}(y_2,y_1)$
using following iterative scheme
\begin{eqnarray}
z^{(2)}_X(y_2) &=& \int  {\mathrm d} y_1 \, a_X(y_2,y_{1}) \nonumber \\
z^{(3)}_X(y_3) &=& \int  {\mathrm d} y_2 \, a_X(y_3,y_{2}) z^{(2)}_X(y_2) 
\nonumber \\
\ldots && \nonumber \\
z^{(N)}_X(y_N)&=&  \int  {\mathrm d} y_{N-1} \, a_X(y_N,y_{N-1}) z_X^{(N-1)}
(y_{N-1}) \nonumber \\
Z_X&=&  \int  {\mathrm d} y_{N} \, z_X^{(N)}(y_{N}). 
\label{sucint}
\end{eqnarray}
\begin{figure}[ht!]
\begin{center}
\includegraphics[width=8cm]{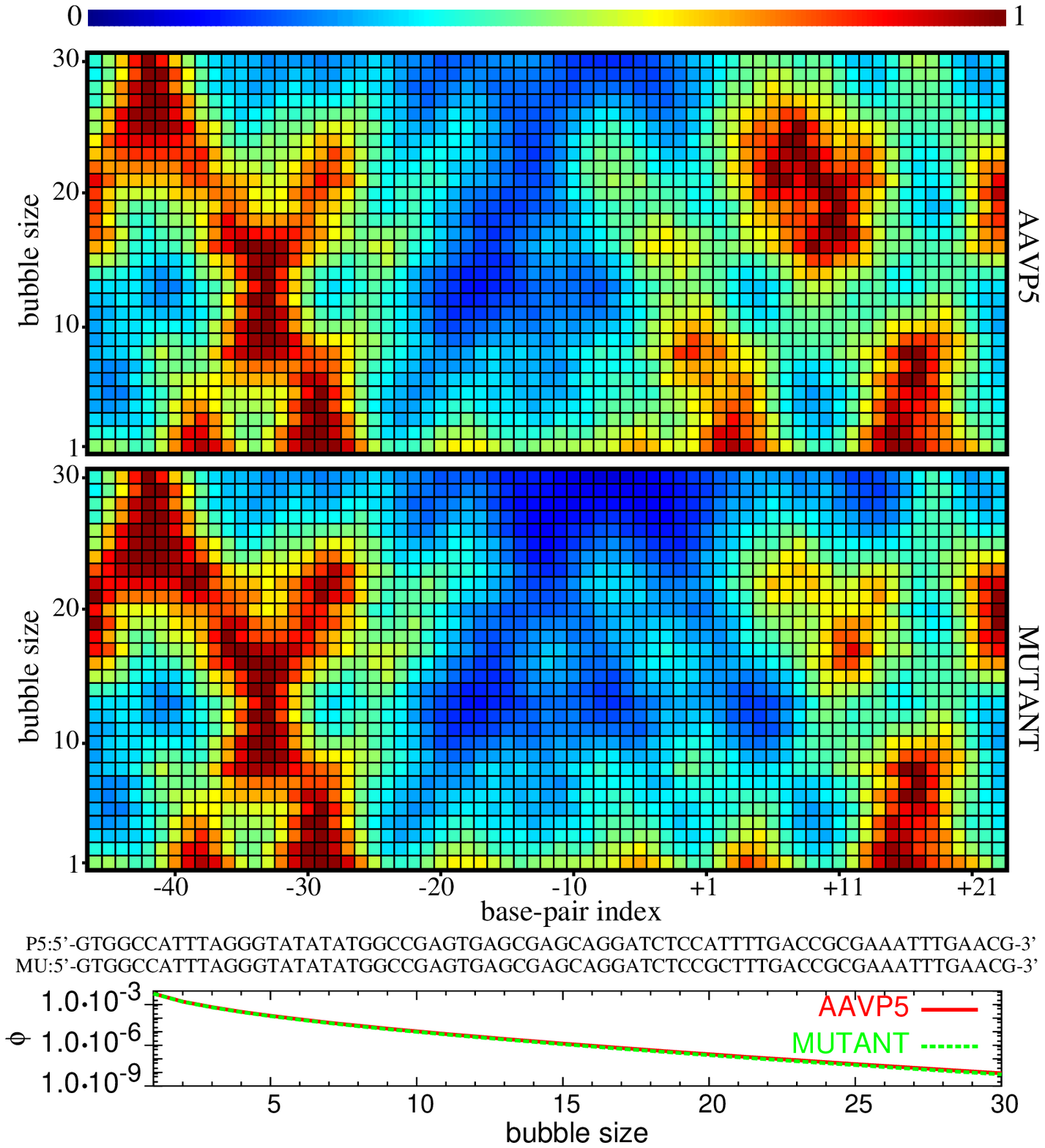}
\end{center}
\caption{(color). The probability of bubble opening as function of bubble
size and position for the
AAVP5 promoter and the mutant sequence at 300 K.
Probabilities in each row are normalized
by a different factor $\phi(m)=\textrm{MAX}[\lo \theta_i^{[m]} \rc_\mu ]
\textrm{ for } i \in [1,N]$
given in the lower panel.
The 69 bp sequences start at index -46 and end at +23.
The TSS is at +1, the mutation is at $(+1,+2)$ were (A,T) bases 
are replaced by (G,C).
Contrary to \cite{ChoiNuc2004}, the mutation effect is very local.}
\label{fig1}
\end{figure}
The calculation of $z_X^{(i)}(y_i)$ for a discrete set of $n_{\rm grid}$
values $y_i$ 
requires only $n_{\rm grid}^2$ function 
evaluations whenever $z_X^{(i-1)}$ is known. Hence, a total 
of $N \cdot n_{\rm grid}^2$ function evaluations are required
instead of $n_{\rm grid}^N$ which is a huge improvement.  
Further increase can be obtained by introducing proper  cut-offs for the 
numerical integration.
We use integration boundaries such that for all  $i$: $L<y_i<R$ and 
$|y_i-y_{i-1}|<d$,
which we control by a single input parameter $\epsilon$: 
$d = \sqrt{\frac{2 |\ln \epsilon|}{\beta K}}
$, $L= -\frac{1}{a_AT}\ln \Big[ \
\sqrt{ \frac{|\ln \epsilon|}{\beta D_{AT}} }+1 \Big]$, and $R=y_0+\sqrt{N}d$.
Any configuration outside this range but with at least one base-pair closed
will have a probability density smaller than $\epsilon/(Z-Z_\Pi)$. 
A strong decrease 
in the parameter $\epsilon$ will only marginally increase the integration 
boundaries. We took $\epsilon=10^{-40}$ that is much smaller than 
necessary for our accuracy.
After storing the following function values in matrices
$M_{ij}^{(AT/GC)}\equiv\exp(-\beta[V_{\rm AT/GC}(L+i \Delta y)+W(L+i\Delta y,
L+(i+j)\Delta y) ])$ with $0 \leq i \leq \textrm{INT}[(R-L)/\Delta y]$ and
$-\textrm{INT}[d/\Delta y] \leq j \leq \textrm{INT}[d/\Delta y]$
we can reduce the integral operations 
for Eq.~(\ref{sucint}) (using Simpson's rule) into
inexpensive multiplication and addition operations 
only.

As a first investigation, we applied this new method  
on the adeno-associated viral P5 promoter and the mutant
from Refs.~\cite{ChoiNuc2004} using $y_0=1.5$ as opening threshold
which corresponds to  2.1 \AA~ in real units.
To make the comparison with MD 
which uses 
periodic 
boundary conditions (PBC), we replicated the chain at both ends, 
but only computed the statistics for the middle chain. This approach,
is cheaper
than true PBC which scales as
$N \cdot (n_{\rm grid})^3$. The full probability matrix
$\lo \theta_i^{[m]} \rc_\mu$ was calculated for the middle sequence
up to bubbles of size $m=50$. A fraction of this matrix is presented
in Fig.~\ref{fig1} in a color plot.
In agreement with Ref.~\cite{ChoiNuc2004}
we find preferential opening probabilities at the 
TSS site at +1 
that vanishes after the mutation. But contrary to the 
results of Ref.~\cite{ChoiNuc2004}, we find that the TSS is not at all
the most dominant opening site. Stronger opening sensitivity is found at the 
-30 region. Moreover at variance with the previous established
findings, Fig.~\ref{fig1} shows that
the mutation effect is very local.
\begin{figure}[ht!]
\begin{center}
\includegraphics[width=8cm]{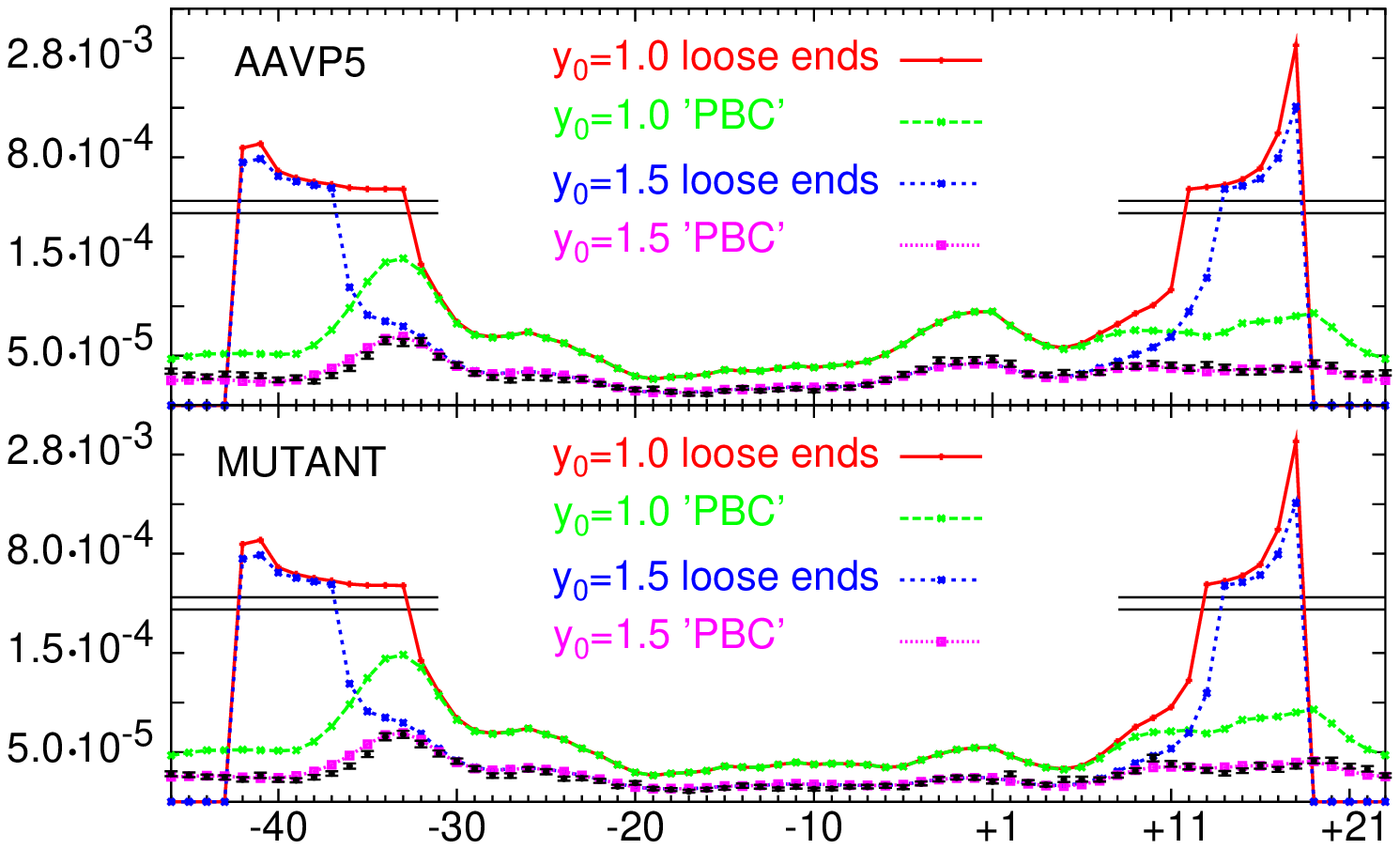}
\end{center}
\caption{(color). The probabilities for bubbles larger than 10 
bp for the AAVP5 promoter and the mutant at 300 K. 
Both semi-PBC (three-fold replicated system) and loose ends (single chain) are
compared and two values for the opening threshold $y_0=1.0$ and $y_0=1.5$.
MD results (black) for $y_0=1.5$ with PBC are also given with 
corresponding error-bars.  A change of scale in the $y$ axis is applied
to include the higher openings at the free boundaries. All results agree but 
are different from the less accurate results of \cite{ChoiNuc2004}. 
The mutation and the free boundaries only have a local impact on the bubble 
statistics.} 
\label{fig2}
\end{figure}
In Fig.~\ref{fig2} we make a projection by looking at the probability
$P_i \equiv \sum_{m=10}^{N-1} \lo \theta_i^{[m]} \rc_\mu$ 
that at site $i$ one can find a bubble of size 10 or larger.
We compared different boundary conditions and two  values for $y_0$.
In addition, we made the comparison with MD~\footnote{
In principle, MD suffers from the same
problem that it allows for complete separation. As a consequence, very
long MD simulations will always give erroneous results.
To restrict the MD to the dsDNA one can use a bias potential
that acts on $y_{\rm min}=\textrm{MIN}[\{y_i\}]$. For instance,
$V^{\rm bias}(y_{\rm min}) = (y_{\rm min}-y_0)^6 \textrm{ if } y_{\rm min} >
 y_0$ and 0 otherwise. However, at 300 K
the complete denaturation occurs so seldom
that it was not detected in all simulations.} by
performing 
100 simulations of 100 ns with different friction constants $\gamma$ 
in the Langevin MD and 10 simulations of 1 $\mu$s 
using Nos\'e-Hoover.
The curves matched within the  
statistical errors and agreed with the integration method (see
for instance Fig.~\ref{fig2} where the Langevin
$\gamma=10$ results are plotted together with the results of the
integration method).

We obtained relative errors around 10 \%
for Nos\'e-Hoover and 
Langevin with $\gamma=10$ and $5$ ps$^{-1}$.
The errors of the $\gamma = 0.05$ ps$^{-1}$, used in Ref.~\cite{ChoiNuc2004},
were considerably larger due a stronger correlation between 
successive timesteps. The results of \cite{ChoiNuc2004} were based 
on 100 times fewer statistics. Hence, the corresponding errors in 
\cite{ChoiNuc2004} 
must have been 
10 times larger which can explain the variance 
with our results. 
Another explanation could be that the results of \cite{ChoiNuc2004}
are due to some out-of-equilibrium or dynamical effects. 
Such effects depend strongly on the choice of initial
conditions, which poses the problem of defining biologically
significant initial conditions and determining, in a meaningful way,
 the relevant time scale
along which the simulations have to be carried to detect such
non-equilibrium phenomena.

The principal error in the new method is mainly 
due to the finite integration steps.
To estimate the accuracy, we compared 
$\Delta y=0.1$ and  $0.05$ with the almost exact results of $\Delta y=0.025$.
Using the TSS peak of  the AAVP5 sequence with free boundaries as reference,  
we found that the systematic error drops 
from $\sim 5$~\% to 0.03~\% for  CPU times of 40 minutes and 3 hours only.
For comparison, the last accuracy would take about 
200 years with MD on the same machine.   
The evaluation of larger bubbles becomes increasingly more difficult for MD.
Bubbles of size 20 showed statistical errors $> 100$ \% while these were
only slightly increased for the integration method. It is interesting to note
that the 10 bp size is 
more or less the upper limit for which one get sufficient accuracy using MD, while
it is a lower limit were its relation to biophysics becomes 
interesting~\cite{Murakamiscience}   
stressing the importance of our method.
\begin{figure}[ht!]
\begin{center}
\includegraphics[width=8cm]{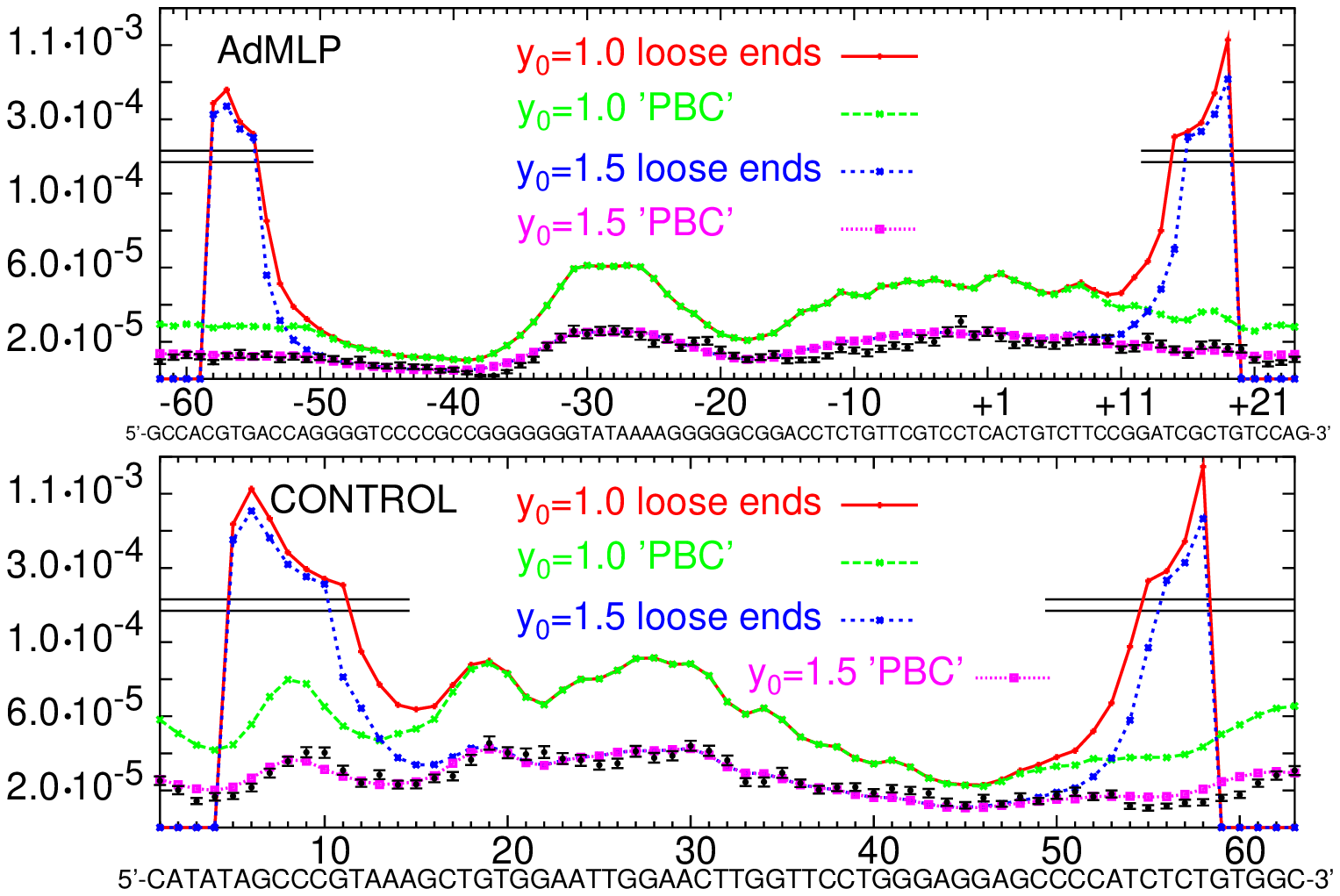}
\end{center}
\caption{(color). Same as Fig.~\ref{fig2} for the 86 bp AdMLP and the 63 bp
non-promoter control sequences. The biological non-active control sequence
shows considerable opening probability, even more than the biological active
AdMLP promoter.}
\label{fig3}
\end{figure}
Finally, we calculated the 
$P_i$ probabilities for the adenovirus major late promoter (AdMLP) and 
a control non promoter sequence (Fig.~\ref{fig3}).
Also here, our results violate the TSS conjecture.
The TSS shows some opening, but cannot be assigned on basis of bubble profile
only.
Surprisingly, even
the control sequence shows significant opening probabilities. 

To conclude, we have shown that MD (or MC) encounters difficulties to
give a precise indication of preferential
opening sites. In particular, information of large bubbles is not easily
accessible using standard methods.
The method presented here is orders of magnitude faster than MD without
imposing additional approximations. 
Using this method, we showed that the TSS is generally not the most
dominant opening site for bubble formation.
These results contradict foregoing conjectures based on less accurate
simulation techniques. 
However, to address the title's question, definitely, there are still many
issues to be solved.
Still, there is some chance that bubble dynamics rather than bubble statics
is indicatory for the TSS. Speculatively, the previously found correlation
could be justified using this argument. However, a statistical significant
foundation for this is lacking and it is highly questionable
whether the PBD model and this type of Langevin dynamics 
can give a sufficiently accurate description for the dynamics 
of DNA.
The PBD model could and, probably, should be improved 
to give a correct representation of 
the subtile sequence specific properties of DNA.
Base specific stacking interaction seems to give
better agreement with some direct experimental observations~\cite{Santiago}.
Also, the development of new experimental techniques is highly desirable.
Our method is not limited to the PBD model or to bubble statistics
only, but it works whenever the
proper factorization~(\ref{sucint}) can be applied. Therefore, we believe
that the technique presented here will remain of 
importance for  the future investigations of bubbles in DNA and their biological
consequences.

We thank Dimitar Angelov and David Dubbeldam for fruitful discussions.
TSvE is supported 
by a Marie Curie Intra-European Fellowships
(MEIF-CT-2003-501976) within
the 6th European Community Framework Programme.
SCL is supported by the Spanish Ministry of Science and
Education (FPU-AP2002-3492), project BFM 2002-00113 DGES and DGA (Spain).

\bibliographystyle{prsty}

\end{document}